\documentclass[reprint %
 reprint,
 amsmath,amssymb,
 aps,
pra,
]{revtex4-2}
\usepackage[T1]{fontenc}
\usepackage{textcomp}
\usepackage{mathptmx}
\usepackage[scaled=.92]{helvet}
\usepackage{courier}
\usepackage{graphicx}
\usepackage{newtxtext}
\usepackage{newtxmath}
\usepackage{natbib}
\usepackage{subcaption}
\usepackage{siunitx}

\usepackage{hyperref}
\hypersetup{
    colorlinks = true,
    urlcolor   = blue,
    citecolor  = black,
}

\newcommand{\RomanNumeralCaps}[1]
\linenumbers

\usepackage{bm}
\newcommand{\vdot}{\cdot}
\newcommand{\na}{\nabla}

\newcommand{\eps}{ \varepsilon}  

\newcommand{\pdif}[2]{ \frac{\partial #1}{\partial #2}}
\newcommand{\dpdif}[2]{ \frac{\partial #1}{\partial #2}}
\newcommand{\Per}{\mathrm{Pe}_r}

\usepackage[T1]{fontenc}
\usepackage{textcomp}
\usepackage{mathptmx}
\usepackage{booktabs} 
\usepackage{graphicx}
\usepackage{bm}
\usepackage{newtxtext}
\usepackage{newtxmath}
\usepackage{hyperref}
\usepackage{cleveref}
\usepackage{enumitem}

\newcommand{\Pe}{P\kern-.06em e}
\usepackage{bm}
\usepackage{letltxmacro}
\LetLtxMacro{\originaleqref}{\eqref}
\renewcommand{\eqref}{Eq.~\originaleqref}
\renewcommand*{\eqref}[1]{Eq.\,\originaleqref{#1}}
\begin{document}
\preprint{APS/123-QED}

\title{Bacterial adhesion to curved surfaces in fluid flow}

\author{Edwina F. Yeo}
 \email{Contact: edwina.yeo.14@ucl.ac.uk}

\affiliation{Department of Mathematics, University College London, London, WC1H 0AY, UK}

\author{Benjamin J. Walker}
\affiliation{Department of Mathematics, University College London, London, WC1H 0AY, UK}

\author{Philip Pearce}
\affiliation{Department of Mathematics, University College London, London, WC1H 0AY, UK}
\affiliation{Institute for the Physics of Living Systems, University College London, London, WC1H 0AY, UK}

\author{Mohit P. Dalwadi}
\affiliation{Mathematical Institute, University of Oxford, Oxford, OX2 6GG, UK}
\affiliation{Department of Mathematics, University College London, London, WC1H 0AY, UK}

\begin{abstract}
Minimising bacterial surface adhesion and subsequent biofilm formation in industrial and medical settings requires understanding how bacteria are transported and adhere to complex surface geometries in the presence of non-uniform flow.  In this paper, we consider the transport of a dilute suspension of motile bacteria through a corrugated two-dimensional channel with perfectly adhesive walls. 
We asymptotically analyse the diffusive boundary layer that forms in high velocity flows using a curvilinear coordinate system based on the fluid streamfunction, presenting a similarity solution to the diffusivity-varying diffusion-type equation that arises.
From this solution, we derive an analytical expression for the bacterial adhesion rate as a function of surface arclength and the spatially varying wall shear rate. Our model predicts that bacterial adhesion becomes localised on curved surfaces, with bacteria showing preferential adhesion to wall `peaks' at lower shear rates and preferential adhesion to wall `valleys' at higher shear rates.  More broadly, our results highlight how spatially varying flows generated by complex geometries can lead to localised bacterial adhesion, with potential implications for both enhancing and minimising biofilm formation. 

\end{abstract}

\maketitle

\section{Introduction}

Surface colonisation by bacteria leads to healthcare-associated infections, the fouling of food products, and the loss of efficacy of wastewater treatment \citep{shirtliff2009role,camara2022economic}. 
Many common pathogens reorient and propel themselves in search of nutrients or surfaces on which they can form a biofilm \citep{wei2011population}.
This motility means bacteria can exhibit more complex dynamics in flow than passive particles. Even in unidirectional shear flow, bacterial surface adhesion rates have been measured in experiments to either increase or decrease as a function of the fluid shear depending on the experimental setup \citep{moreira2014effects,lecuyer2011shear,palalay2023shear,saur2017impact,park2011effect}. Recent theoretical analysis of bacterial adhesion in shear flow has predicted a nonmonotonic adhesion rate as a function of shear rate in simple channel flow, demonstrating that adhesion is reduced by upstream bacterial reorientation at high flow rates \citep{YeoPNAS}.  
However, many medical and industrial systems are associated with more complex geometries with textured surfaces, bends and constrictions. The effects of these geometries on bacterial adhesion have not been theoretically quantified.

The lengthscale of constrictions, obstacles or textured surfaces relative to the bacterial body size is a key determinant in predicting their effect on bacterial adhesion \citep{cheng2019micro}. Surface modifications of lengthscales below \SI{5}{\micro\metre} can reduce bacterial attachment by actively killing bacteria on contact~\citep{tripathy2017natural}, or by limiting the accessible area available for bacteria to adhere \citep{pellegrino2026reduction}. 
Large lengthscale modifications ($>$\SI{5}{\micro\metre}) generate regions of high and low fluid shear, significantly altering the hydrodynamic transport of bacteria. In experiments, this has been found to have different effects on bacterial surface accumulation depending on the size of the wall shear rate compared to the bacterial swimming speed: bacteria accumulate downstream of obstacles or constrictions for relatively low shear rates of $<$\SI{30}{\per\second} \citep{Secchi2020, Mino2018,altshuler2013flow} and upstream of obstacles at relatively high shear rates of $\approx$\SI{200}{\per\second} \citep{Secchi2020}; this transition has been replicated in agent-based models \citep{Secchi2020, lohrmann2023novel,lee2021influence}. The transition from downstream to upstream adhesion has been attributed to the reduced effect of bacterial motility as the flow speed increases above the bacterial swimming speed; for passive particles, adhesion would be expected to be higher upstream. For even higher shear rates $\approx$\SI{1000}{\per\second}, bacteria have been found to accumulate at the bottom of valleys or cavities, potentially because of increased fluid erosion at peaks  \citep{scheuerman1998effects,lee2013flow}. These experimental findings demonstrate that it is challenging to disentangle the competing surface effects of bacterial attachment, boundary shape, fluid erosion and growth in experiments. Therefore, in this paper we will use a theoretical approach to isolate the roles of hydrodynamics and motility in facilitating bacterial attachment to curved surfaces, focusing our analysis on the high shear environments typical of industrial and medical settings.

The effect of high speed flow on the adhesion of passive particles in straight and curved channels has previously been quantified theoretically. Classical work predicts that increasing flow increases mass transport and, therefore, adhesion; a boundary layer analysis in shear flow yields a surface adhesion rate of $J\sim\dot\gamma^{1/3}D^{2/3}x^{-1/3}$ where $D$ is the particle diffusion coefficient, $\dot \gamma$ is the wall shear rate and $x$ is the distance from the particle source \citep{leveque1928lois}. An extension of this result to more complex geometries was provided by \cite{lighthill1950contributions}, who analysed thermal boundary layers over general surfaces using an arclength rescaling method.
More recently, we theoretically quantified the effects of activity on adhesion in straight channels, 
predicting that the adhesion rate of motile bacteria with swimming speed $\mathcal V_s$ and rotational diffusion coefficient $D_r$ 
in shear flow depends nonmonotonically on the wall shear rate \citep{YeoPNAS}. This is captured by an effective diffusivity $D_{\mathrm{eff}}$. At low shear, $D_{\mathrm{eff}}\sim\mathcal{V}_s^2/2D_r$, in agreement with the quiescent approximation of \cite{berg1993random}, giving $J\sim\dot\gamma^{1/3}$; at high shear, upstream bacterial reorientation limits diffusivity and, therefore, adhesion, with $J\sim\dot\gamma^{-1}$ instead \citep{YeoPNAS}.  Generalising these results to understand active particle adhesion to curved surfaces 
requires characterising the effective diffusivity of bacterial suspensions in high-speed, spatially varying fluid flows near boundaries.  

The diffusivity of bacterial suspensions in the absence of adhesion has predominantly been estimated using generalised Taylor dispersion (GTD) in systems including unidirectional flows \citep{frankel1991generalized,bearon2003extension}, quiescent suspensions in slowly varying corrugated channels \citep{yariv2014ratcheting} and porous media flow \citep{alonso2019transport}.  An alternative approach to GTD was used in \citet{Fung2022}, which approximates bacterial dispersion in multidirectional flows using the local flow gradient. However, the large gradients in both density and orientation that form in higher velocity flows mean that the requisite assumptions of spatial homogeneity used in both GTD and the local approximation model of \citet{Fung2022} cannot be used.  Hence, the combined effect of high velocity flow and curved boundaries on bacterial adhesion remains to be quantified theoretically.

In this paper, we consider the perfect adhesion of bacteria to corrugated surfaces in flow (\S\ref{section:problem}). Using a curvilinear coordinate system defined via the fluid streamfunction (\S\ref{section:sline-def}), we asymptotically analyse the boundary layer that forms at high flow rates (\S\ref{section:asymptotics}). We then derive a similarity solution to the subsequent boundary layer equation using an adaptation of the flow approximation and arclength rescaling in \cite{lighthill1950contributions}.  Using this similarity solution, we calculate an analytic expression for the bacterial adhesion rate in terms of the local wall shear rate (\S\ref{section:bdy-layer}). We evaluate this solution using numerical Stokes 
flow solutions, demonstrating that bacterial adhesion can be localised to either wall `valleys' or `peaks' depending on the ratio of the bacterial rotational diffusion to the fluid shear rate (\S\ref{section:results}). By varying the surface geometry we show that bacterial localisation is further exaggerated when the curved walls have tall obstacles with short wavelengths, which most significantly disturb the flow. 
\label{sec:headings}
\begin{figure}
\centering
\includegraphics[width=0.7\textwidth]{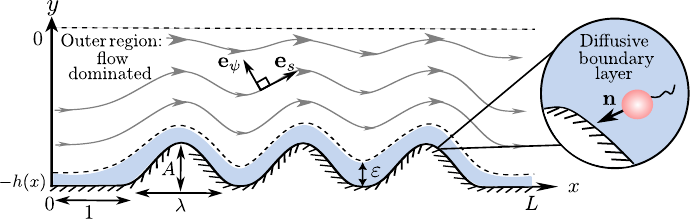}
  \caption{Schematic for flow of a bacterial suspension through a channel with corrugated, absorbing walls. The asymptotic structure is split between the flow-dominated outer region and the diffusive boundary layer near $y=-h(x)$ (shaded blue) with thickness $\varepsilon$, where bacteria are mainly oriented towards the boundary (inset). Lower half of domain $y\in[-h(x),0]$ shown, with symmetry across $y=0$. Also shown are: representative fluid streamlines (grey), arclength and streamfunction unit vectors $(\bm{e}_s,\bm{e}_\psi)$ and wall geometry amplitude $A$ and wavelength $\lambda$.  }
\label{fig:schematic}
\end{figure}

\section{Problem statement}\label{section:problem}
We consider the steady flow of a dilute suspension of bacteria through a two-dimensional channel defined by $|y|<h(x)$, $0<x<L$, with corrugated, perfectly absorbing walls at $|y|=h(x)$, as shown in Fig.\,\ref{fig:schematic}. Symmetry allows for consideration of only $y \in(-h(x),0)$. We take the bacteria to be circular rigid bodies that experience rotational diffusion with coefficient $D_r$ and propel themselves with fixed speed $\mathcal{V}_s$ in the direction they are facing. We consider circular bacteria since the role of bacterial shape was found to only have a small effect on adhesion at high shear by \cite{YeoPNAS}.  We use the dilute active suspension model presented by \citet{saintillan2015theory}, additionally assuming that the suspension is dilute enough to neglect physical and hydrodynamic interactions between bacteria; the latter assumption implies that the fluid flow is independent of the bacteria. 

We present the governing equations in dimensionless form, obtained by the following scalings: length with the maximum half-channel width $d$; fluid velocity with the maximum inlet velocity $U_{\text{max}}/2$; fluid pressure with the viscous pressure scale $\mu U_{max}/2d$; and bacterial density with the (constant) inlet density $\rho_{i}$.
Under these scalings, the steady Stokes equations apply
for the dimensionless fluid velocity field $\bm{u}= u\boldsymbol{e}_x + v \boldsymbol{e}_y$ (where $\boldsymbol{e}_i$ is the unit vector in the $i$ direction) and pressure $p$:
\begin{align}
    \na^2\bm{u}=\na p, \quad \na\cdot \bm{u}=0.\label{stokes}
\end{align}
The dimensionless bacterial density $\rho$, mean orientation vector $\bm{n}$ and nematic order tensor $\bm{Q}$ are governed by the dilute active fluid equations \citep{saintillan2015theory}
\begin{align}    
   \bm{u}\cdot \na\rho=&-V_s\na\cdot(\rho\bm{n}),\label{rho_eqn}\\
             \bm{u}\cdot \na(\rho\bm{n})=&-V_s\left(\na\cdot(\rho\bm{Q})+\frac{1}{2}\na\rho\right) +(\rho\bm{I} \bm{n}-\bm{T}):\bm{W}-\frac{\rho\bm{n}}{\Per},\label{n_eqn}
\\
 \bm{u}\cdot \na(\rho\bm{Q})=&-V_s\left(\na\cdot\bm{T}+\frac{\bm{I}}{2}\na\cdot(\rho\bm{n})\right)
         +\rho(\bm{W}\bm{Q}-\bm{Q}\bm{W})-\frac{4\rho\bm{Q}}{\Per}.\label{q_eqn}
           \end{align}
Equations \eqref{n_eqn}-\eqref{q_eqn} capture advection, bacterial swimming, and rotation of bacteria by the flow which is proportional to the rate of rotation tensor $\bm{W}:=(\na\bm{u}-\na{\bm{u}}^T)/2$.
Two dimensionless parameters define the bacterial dynamics: $V_s=2\mathcal{V}_s/U_{\text{max}}$, the ratio of the bacterial swimming speed to the flow speed, and $\Per=U_{\text{max}}/2dD_r$, the rotational P\'eclet number, which captures the ratio of fluid rotation to reorientation by diffusion.
Equations \eqref{rho_eqn}-\eqref{q_eqn} are obtained through successive moments of a Fokker-Planck equation describing the dynamics of a collection of stochastic bacteria. To close this system, higher-order moments must be approximated through the use of a closure. We use the two-dimensional isotropic closure, analogous to a Fourier series in orientation in 2D, which defines components of the rank three tensor in \eqref{q_eqn} as $ 
\bm{T}_{ijk}=\rho(\delta_{ij}n_{k}+\delta_{ik}n_{j}+\delta_{jk}n_i)/4$.

At the inlet ($ x=0$) we impose a fixed  bacterial density $\rho=1$, unbiased swimming direction $\bm{n}=\bm{0}$, uniform angular distribution $\bm{Q}=\bm{0}$, and a unidirectional, parabolic fluid velocity profile $(u,v)=(2(1-y^2),0)$. The prefactor of two in the inlet velocity means that $u\sim y$ near the wall. On the lower channel wall ($y=-h(x))$ we impose perfect bacterial absorption $\rho=0$, along with no slip and no flux $\bm{u}=\bm{0}$.  At the channel outlet ($x=L$) we impose no diffusive flux of bacterial density along with unidirectional flow $v=0$ and atmospheric pressure $p=0$. At the channel centreline ($y=0$) we impose symmetry conditions: $v=0,\  \partial \rho/\partial y=0,\ \partial u/\partial y=0. $

 In the simulations, the corrugated channel walls are defined by an initial straight section, then three peaks of variable amplitude $A$ and wavelength $\lambda$ and a subsequent straight section, which allows the flow to relax back to a unidirectional profile at the outlet located at $x=L$, with $h(x)=1-A\sin^2(\pi(x-1)/\lambda)$ for $1<x<3\lambda+1$ and $h(x)=1$ otherwise.

\subsection{Arclength-streamfunction coordinate system}\label{section:sline-def} 
We analyse  \eqref{rho_eqn}-\eqref{q_eqn} in curvilinear (arclength-streamfunction) coordinates
$(s,\psi)$ (as used previously in, e.g., \citet{batchelor1956steady,cummings2007tissue,booth2024optimal}). Here, $s$ describes the arclength along each streamline, with $s=0$ at $x=0$, and the fluid streamfunction $\psi$ is defined by $u = \partial \psi / \partial y$, $v = - \partial \psi / \partial x$.  This coordinate system has unit vectors $\bm{e}_s$ and $\bm{e}_\psi$, which lie tangent and normal to fluid streamlines, respectively (shown in Fig.\,\ref{fig:schematic}). By definition, the fluid velocity is purely in the $\bm{e}_s$ direction, with variable magnitude  $|\bm{u}|=U(s,\psi)>0$ such that 
    $\bm{u}=U(s,\psi)\bm{e}_s
$. In this coordinate system, the gradient and divergence operators are
\begin{align}
    \na f&=\pdif{f}s\bm{e}_s+U\pdif{f}\psi\bm{e}_\psi,\quad
    \na \vdot\bm{g}=U\left(\pdif{}s\left(\frac{g_s}{U}\right)+\pdif{(g_\psi)}\psi\right)
\end{align}
for a scalar function $f$ and a vector function $\bm{g}=g_s\bm{e}_s+g_\psi\bm{e}_\psi$,
where derivatives with respect to $\psi$ capture shear gradients, and derivatives with respect to $s$ capture flow acceleration along a streamline. We consider corrugated surfaces of moderate curvature for which there are no regions of recirculation, allowing us to uniquely set $\psi=0$ on $y=-h(x)$.

\section{Analytical and numerical solution}\label{section:asymptotics}
We consider the regime in which fluid flow is much faster than the swimming speed of an individual bacterium, corresponding to $V_s\ll1$, in the distinguished limit in which $\Per=O(1)$. In this regime, the asymptotic structure of the solution is divided into an outer region where bacteria advect as passive tracers, and an inner (boundary) layer where complex bacterial dynamics and adhesion alter bacterial density. Conceptually, this structure is similar to the problem of bacterial adhesion past a flat plate \citep{YeoPNAS}, though here with curved boundaries and multidirectional flow.

\subsection{Flow-dominated outer region}
In the weak-swimming limit $V_s\ll1$, far from the absorbing boundary at $y=-h(x)$, \eqref{rho_eqn} states that bacteria remain on fluid streamlines at leading order, with a constant density $\rho=1$ set by the inlet condition. 
Then, the leading-order versions of \eqref{n_eqn} and \eqref{q_eqn} consist of a balance between rotational diffusion, fluid rotation and advection along streamlines: 
\begin{align}    
            U\pdif{(\rho\bm{n})}s=(\rho\bm{I} \bm{n}-\bm{T}):\bm{W}-\frac{\rho\bm{n}}{\Per},\qquad
      U\pdif{(\rho\bm{Q})}s=
            \rho(\bm{W}\bm{Q}-\bm{Q}\bm{W})-\frac{4\rho\bm{Q}}{\Per}.\label{q_eqnbdy}
           \end{align}
Using the homogeneous inlet conditions, these generate the solutions $\bm{n}=\bm{Q}=\bm{0}$. That is, in the outer region, the bacteria are uniformly oriented with unbiased swimming direction.

\subsection{Boundary layer}\label{section:bdy-layer}
Close to the curved wall, no slip means that the fluid flow slows and eventually becomes comparable to the magnitude of bacterial swimming. When this occurs, a diffusive boundary layer forms where bacterial density is determined by a balance of streamwise advection and swimming normal to the streamlines. The thickness of this layer, $\eps$, will be identified as part of our subsequent analysis. Since $u \sim y$ near the wall, the flow in this boundary layer is of $O(\eps)$ and, hence, the streamfunction is of $O(\eps^2)$. We consider surfaces with moderate curvature $|\kappa|\ll1/\eps$. We scale into this boundary layer region by defining a new streamfunction $\psi=\eps^2\Psi$ and fluid velocity $\bm{u}=\eps\tilde U\bm{e}_s$, denoting other dependent variables with tildes. The leading-order scaled version of \eqref{rho_eqn} is
\begin{align}
  \eps \tilde{U}\pdif{\tilde \rho}s+\frac{V_s  \tilde{U}}{\eps}\pdif{(\tilde\rho \tilde n_\Psi)}\Psi=0\label{rho-bdy1}.
\end{align}
Here, $\tilde \rho \tilde n_\Psi$ captures the proportion of bacteria oriented normal to streamlines. We seek solutions such that swimming balances advection in \eqref{rho-bdy1}, so that $| \tilde {\bm{n}}|\ll1$. Rescaling \eqref{n_eqn}, this is achieved by a balance between fluid rotation, rotational diffusion, and swimming of bacteria down density gradients in the $\Psi$-direction, resulting in the leading-order equations
\begin{align}  
  0&=\frac{1}{2}\tilde U \pdif{\tilde U}\Psi \tilde \rho \tilde n_\Psi-\frac{\tilde \rho\tilde  n_s}{\Per},\quad      0=-\frac{1}{2}\tilde U \pdif{\tilde U}\Psi\tilde\rho\tilde n_s-\frac{\tilde \rho\tilde  n_\Psi}{\Per} -\frac{V_s\tilde U}{2\eps}\pdif{ \tilde \rho}{\Psi}.\label{nb2_eqn0} 
\end{align}
Here, we only see shear gradients of the flow, which are dominant in the boundary layer. 
We can rewrite \eqref{nb2_eqn0} to determine the components of $\rho \tilde  {\bm{n}}$ in terms of the other variables:
\begin{align}
\tilde  \rho \tilde  n_\Psi&=-\frac{2V_s\Per\tilde{U}}{\varepsilon\left(  \Per^2\tilde U^2\left(\dpdif{\tilde U}\Psi\right)^2+4\right)}\dpdif{\tilde \rho}\Psi, \label{polar-bdy} \quad   \tilde \rho \tilde n_s=-\frac{V_s\Per^2\tilde{U}\left(\dpdif{\tilde U}\Psi\right)^2}{\varepsilon\left(  \Per^2\tilde U^2\left(\dpdif{\tilde U}\Psi\right)^2+4\right)}\dpdif{\tilde \rho}\Psi. 
\end{align}
Hence, the mean swimming direction $\tilde{\bm{n}}$ points down gradients in bacterial density perpendicular to the flow, with magnitude that depends on the local fluid shear. The self-consistent asymptotic scalings suggest that the nematic order tensor is much smaller than the retained terms, so it does not appear in \eqref{polar-bdy}. Combining \eqref{rho-bdy1} and \eqref{polar-bdy}, we obtain a closed (spatially varying diffusion) equation for the leading-order bacterial density
\begin{align}\pdif{\tilde \rho}s+\frac{2V_s^2\Per}{\eps^3}\pdif{}\Psi\Bigg(\frac{\tilde{U}}{  \Per^2\tilde U^2\left(\dpdif{\tilde U}\Psi\right)^2+4}\dpdif{\tilde \rho}\Psi \Bigg)=0,\label{lo-bact}
\end{align} with boundary condition $\tilde\rho=0$ on $\Psi=0$ and matching condition $\tilde\rho\to1$ as $\Psi \to \infty$. It is convenient to define the boundary layer thickness by considering the scale of $\tilde U$ and $\partial\tilde U/\partial \Psi$ in the straight portion of the boundary, giving  $\varepsilon^3=2V_s^2\Per/(\Per^2+4)$.

Following \citet{lighthill1950contributions,fage1931further} for passive particles, Equation \eqref{lo-bact} admits a similarity solution in the scenario where the boundary layer flow is well approximated by shear flow with dimensionless shear rate $\dot\gamma(s)$.
In this case, the boundary layer flow can be defined as $\tilde U=\sqrt{2\dot\gamma(s)\Psi}+O(\eps)$\footnote{This reduces to unidirectional shear flow ($\tilde U=\sqrt{2\Psi}$) when the surface is flat and $\dot \gamma(s)=1$.}, 
and \eqref{lo-bact} becomes \begin{align}
\pdif{\tilde \rho}s+\frac{\sqrt{\dot\gamma(s)}\left(  \Per^2+4\right)}{\left(  \Per^2\dot \gamma(s)^2+4\right)}\pdif{}\Psi\left(\sqrt{2\Psi}\pdif{\tilde\rho}\Psi \right)=0.\label{lo-bact2}
\end{align}
Using a nonlinear scaling of the arclength $\xi(s)$, we can transform \eqref{lo-bact2} into
\begin{align}
\pdif{\tilde \rho}\xi+\pdif{}\Psi\left(\sqrt{2\Psi}\pdif{\tilde \rho}\Psi \right)=0\quad \text{with }\quad   \xi=\int_0^s \frac{\sqrt{\dot\gamma(\hat s)}\left(  \Per^2+4\right)}{\left(  \Per^2\dot \gamma(\hat s)^2+4\right)}d\hat s.\label{lo-bact3}
\end{align}
Then, we obtain a similarity solution to \eqref{lo-bact3} using the similarity variable $\eta=\Psi \xi^{-2/3}$. Imposing the appropriate boundary and matching conditions, we obtain the solution 
\begin{align}
     \rho(s,\psi)=\frac{1}{\Gamma(1/3)}\Gamma\left(\frac{1}{3},\frac{4 \,\psi^{3/2}}{9 \sqrt{2}V_s^2\Per\int_0^s \sqrt{\dot\gamma(\hat s)}\left(  \Per^2\dot \gamma(\hat s)^2+4\right)^{-1}  d\hat s}\right), \label{solution}
 \end{align} which we write in terms of the standard ($s,\psi)$ coordinates. In \eqref{solution}, $\Gamma(a,z)$ is the lower incomplete gamma function and $\Gamma(a)$ is the gamma function. 
 
 The pointwise bacterial surface adhesion rate $J(s)$ is defined by the vertical swimming flux of bacteria, $J(s)=-V_s\rho n_\psi|_{\psi=0}$. 
 Using \eqref{polar-bdy} and our analytic solution \eqref{solution}, we may derive an analytic expression for the bacterial surface adhesion rate along the curved surface:

\begin{align}
 J(s)&=\frac{2V_s\Per \sqrt{2\dot\gamma(s)\psi}}{\left(  \Per^2\dot\gamma(s)^2+4\right)}\dpdif{\rho}\psi\bigg|_{\psi=0} = \frac{3^{1/3}}{\Gamma(1/3)}\frac{(2V_s^2\Per)^{2/3} \sqrt{\dot\gamma(s)}}{\left(  \Per^2\dot\gamma(s)^2+4\right)}\left(\int_0^s \frac{\sqrt{\dot\gamma(\hat s)}}{\left(  \Per^2\dot \gamma(\hat s)^2+4\right)}  d\hat s\right)^{-1/3}\label{flux}
 \end{align}
  Reassuringly, \eqref{flux} reduces to the flat-plate adhesion result of \cite{YeoPNAS} in the appropriate limit of $\dot \gamma(s)=1$, under the substitutions $s=x$ and $\psi=y^2/2$.

\subsection{Numerical method}
We solve the fluid problem \eqref{stokes} with the Finite Element Method implemented using the Python Package FEniCS \citep{LoggWells2010,LoggEtal_10_2012},
which allows implementation of the weak form in the language UFL \citep{LoggEtal_11_2012,KirbyLogg2006,OlgaardWells2010}. The problem is then compiled by FIAT \citep{kirby2010}. We use GMSH to construct a mesh of the channel geometry \citep{geuzaine2009gmsh}. We then solve for the wall shear stress by computing gradients of the fluid solution.

\section{Results and discussion}\label{section:results}

\begin{figure}
    \centering
    \includegraphics[width=0.8\linewidth]{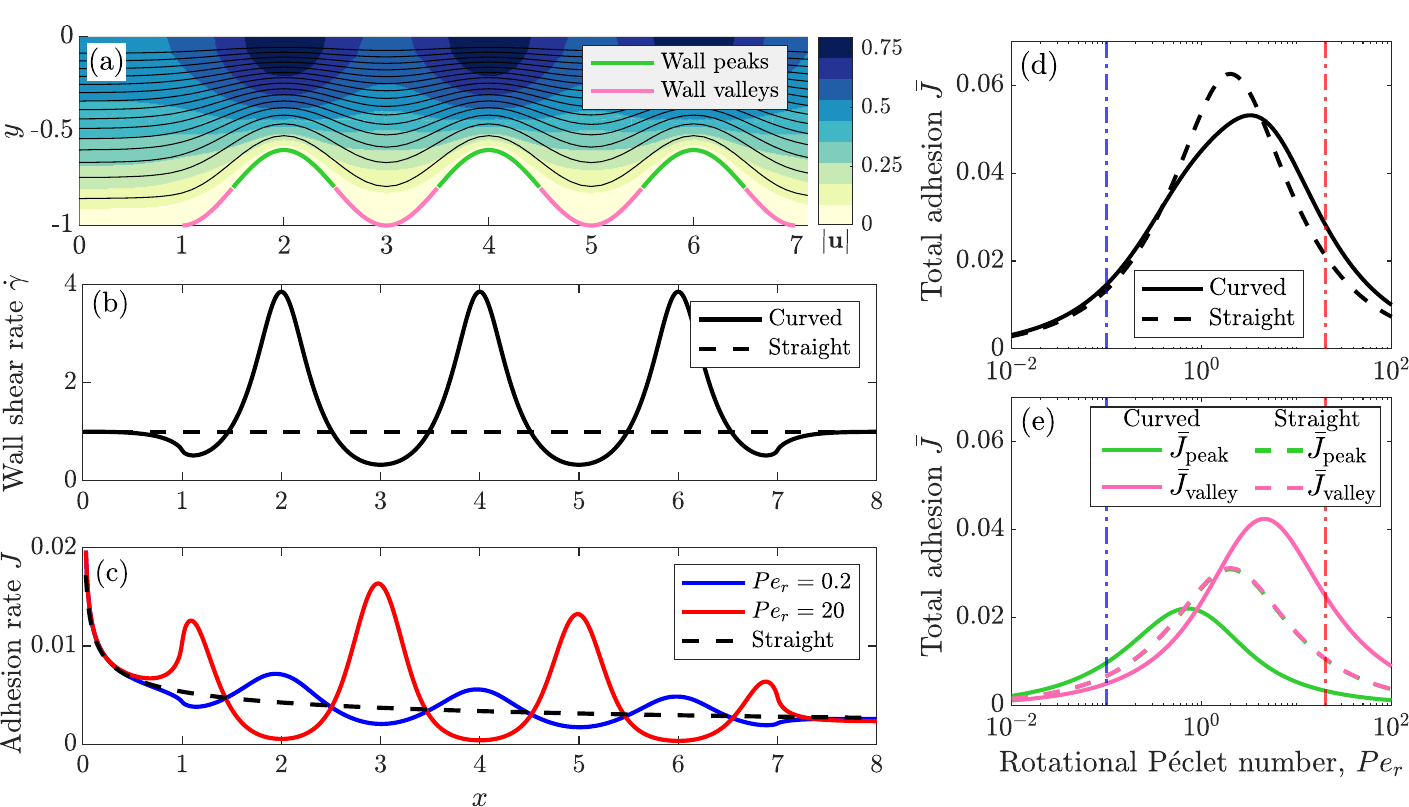}
    \caption{ (a) Fluid velocity magnitude in corrugated channel with black streamlines overlaid; coloured lines mark wall valleys and peaks. (b) Wall shear rate in corrugated and straight channels. (c) Spatial adhesion distribution for $\Per=0.2$ and $\Per=20$.     (d) Total adhesion in corrugated region for a range of $\Per$. (e)   Total adhesion to peaks and valleys for a range of $\Per$ (see panel (a) for relevant parts of the boundary). All results are for $V_s=0.1$ and wall parameters $A=0.4$, $\lambda=2$, $L=12$, with straight channel alternatives shown dashed in (b-e). Vertical lines in (d-e) correspond to $\Per=0.2$ \& $20$ as in (c). }
    \label{fig2}
\end{figure}

We have derived an analytical expression to predict bacterial adhesion to curved surfaces, which requires only the local wall shear rate and the bacterial motility parameters to evaluate via Eq.\,\eqref{flux}. The integral in Eq.\,\eqref{flux} describes the cumulative effect of upstream shear and adhesion on the bacterial density, and the remaining terms describe the instantaneous response of the mean bacterial orientation to local shear rate $\dot \gamma(s)$. To evaluate Eq.~\eqref{flux}, we numerically simulate a spatially varying flow through a corrugated channel, for which the wall shear rate is higher at the wall peaks and lower in wall valleys (Fig.\,\ref{fig2}a,b). This produces spatially localised bacterial adhesion $J(s)$, with bacteria preferentially adhering to wall peaks at low $\Per$ and to wall valleys at high $\Per$ (Fig.~\ref{fig2}c).

A comparison of our corrugated-channel results with straight channels shows that the total adhesion of bacteria $\bar J=\int J dx$ in the corrugated region ($1<x<1+3\lambda$) can be either reduced or increased by the corrugations (Fig~\ref{fig2}d). At intermediate P\'eclet numbers ($\Per\approx1)$ overall adhesion is reduced by up to 20\%;  at lower and higher $\Per$, overall adhesion is increased by up to 35\% (Fig.~\ref{fig2}d). Comparing total adhesion to peaks ($\bar J_\mathrm{peak}$) and to valleys ($\bar J_\mathrm{valley}$) further demonstrates localised adhesion (Fig.\,\ref{fig2}e). This is in contrast to straight channels, for which adhesion varies equally with $\Per$ at all locations along the wall.

\begin{figure}
    \centering

    \includegraphics[width=0.75\linewidth]{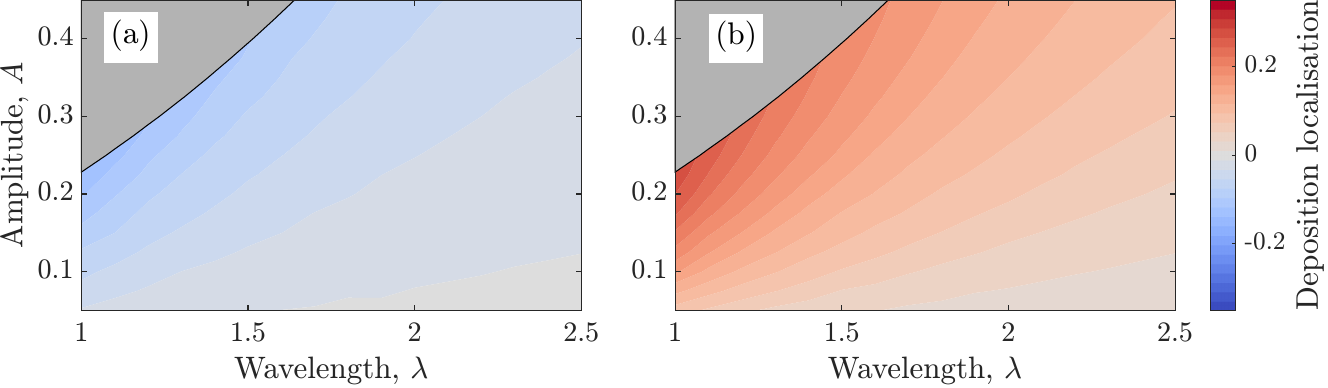}
    
    \caption{Deposition localisation, quantified by $(\bar J_\mathrm{valley}-\bar J_\mathrm{peak})/(3\bar J\lambda)$ for a range of wall parameters with $V_s=0.1$.  (a) Negative values show peak localisation at $\Per=0.2$. (b) Positive values show valley localisation at $\Per=20$. In both panels the strongest localisation occurs for tall, small-wavelength corrugations. In the grey region, flow in valleys is insufficient to apply asymptotic results (see \S \ref{section:results}).}
    \label{fig3}
\end{figure}

Varying the wavelength $\lambda$ and amplitude $A$ of the corrugated walls allows for control of the wall shear rate difference between peaks and valleys. Broadly, larger amplitudes generate greater wall shear rates on the peaks, and smaller wavelengths generate lower wall shear rates in the valleys. The relative localisation per wall wavelength can be quantified by $(\bar J_\mathrm{valley}-\bar J_\mathrm{peak})/(3\bar J\lambda)$, which is positive for valley-localised adhesion and negative for peak-localised adhesion (Fig.\,\ref{fig3}).  
We do not consider walls with sufficiently large amplitudes or small wavelengths for which the flow in the valleys becomes slow and eddies are generated \citep{moffatt1964viscous};  both our assumption of $V_s\ll1$ and our streamfunction formulation break down in this regime, marked in grey in Fig.\,\ref{fig3}.  
The adhesion is peak-localised at low $\Per$ (Fig.\,\ref{fig3}a), and valley-localised at high $\Per$ (Fig.\,\ref{fig3}b), though the precise extent of localisation varies with the corrugation parameters. The most extreme localisation occurs for corrugation with tall amplitude and small wavelength, for which the disturbance to the flow is greatest. 

We can mechanistically understand the localised adhesion by examining Eq.\,\eqref{flux} at low and high rotational P\'eclet numbers:
\begin{align}J(s)\sim
\begin{cases}\frac{3^{1/3}}{\Gamma(1/3)}\left(\frac{V_s^2\Per}{2}\right)^{2/3} \dot\gamma(s)^{1/2}
        \left(\int_0^s  \dot\gamma(\hat s)^{1/2} d\hat s\right)^{-1/3}\text{ for } \Per\ll1, \\
           \frac{3^{1/3}}{\Gamma(1/3)} \left(\frac{2V_s^2}{\Per}\right)^{2/3} \dot\gamma(s)^{-3/2}\left(\int_0^s \dot\gamma(\hat s)^{-3/2}  d\hat s\right)^{-1/3} \text{ for }1\ll\Per\ll\eps^{-1}.\end{cases}
        \label{flux:cases}
\end{align}
Although the adhesion rate $J(s)$ has a complex dependence on the upstream shear rate $\dot\gamma(s)$ in \eqref{flux:cases}, the local term indicates that adhesion should broadly be expected to increase with local shear rate at low $\Per$ and decrease with local shear rate at high $\Per$, as is seen in Fig.~\ref{fig2}.
 Physically, this is driven by the instantaneous response of the mean swimming strength $\bm{n}$ to the local wall shear rate, 
 with local increases in flow leading to increased mean swimming speed at low $\Per$ but to upstream reorientation at high $\Per$.  Furthermore, the bacterial adhesion rate responds more strongly to changes in the wall shear rate at high $\Per$, noting the larger exponent of $\dot\gamma(s)$ in \eqref{flux:cases}. This generates a greater disparity between $\bar J_\mathrm{valley}$ and $\bar J_\mathrm{peak}$ at high $\Per$, with up to eight times as much adhesion in valleys. As expected, the low-$\Per$ limit in \eqref{flux:cases} has the same functional dependence on $\dot\gamma$ as the heat flux solution 
 in \cite{lighthill1950contributions}, since the bacteria adhere to the surface like passive particles in this limit. 

Our findings suggest that adding corrugation to straight devices can control initial bacterial adhesion locations and reduce total adhesion at certain flow rates. Furthermore, our results predict that the initial attachment locations of bacterial species with differing motility parameters could be separated, with some attaching to low-shear valleys and others to high-shear peaks.  In practice, overall surface colonisation is the result of both attachment and detachment. Fluid flow has been shown to both increase bacterial detachment, as shear forces overcome attachment \citep{busscher2006microbial}, but also to decrease attachment for specific bacterial surface bonds (FimH in \textit{E. coli} \citep{sauer2016catch}) and cell shapes (\textit{P. aeruginosa} \citep{sauer2016catch}). This difference in post-attachment bacterial response to fluid flow could allow for further control of the biofilm formation process. For example, if shear-enhanced detachment is combined with peak localisation, then overall colonisation would be reduced. However, if shear-enhanced detachment is combined with valley localisation, this could create sheltered zones for bacteria to grow.

Our analysis could be extended to consider elongated bacteria, although the generalized version of \eqref{lo-bact} would feature additional advection terms generated by a non-zero nematic order tensor, so the self-similar scaling could not applied directly. Scenarios where bacteria adhere imperfectly could be considered by imposing a boundary condition on the mean orientation vector on the walls. To ensure this is well posed, Brownian diffusion is often added to such systems \eqref{rho_eqn}-\eqref{q_eqn} (see \cite{maretvadakethope2023interplay}); here, this may result in additional boundary layers forming. 
The analytical bacterial adhesion rate \eqref{flux} relies on our assumption that the flow is well approximated by local shear flow in the diffusive boundary layer. This can hold for flows with non-zero inertia, as long as the viscous boundary layer is much larger than the diffusive one. Similarly, our approach would have to be adapted for systems with more extreme curvature than considered here. 

In summary, this work quantifies the role of curved geometry on bacterial surface attachment in high velocity flows,  demonstrating that nonuniform surface shear can induce localised bacterial adhesion at regions of both higher or lower shear, depending on the scenario of interest. Our focus on high speed flows allows this work to have implications for applications including urinary catheters, food processing facilities and waste water treatment. 

\subsection*{Acknowledgments and Funding}
{E.F.Y is supported by an EPSRC National Fellowship in Fluid Dynamics [EP/X027902/1]. P.P. is supported by a UKRI Future Leaders Fellowship [MR/V022385/1].}

\bibliography{lib}

\end{document}